\documentclass[preprint, 12pt]{aastex}
\usepackage{graphicx,natbib,amsmath,multirow}

\shorttitle{Magnetic Quenching of Turbulent Diffusivity: Reconciling Mixing-length Theory Estimates with Kinematic Dynamo Models of the Solar Cycle} \shortauthors{Mu\~noz-Jaramillo,
Nandy and Martens}

\begin{document}

\title{Magnetic Quenching of Turbulent Diffusivity: Reconciling Mixing-length Theory Estimates with Kinematic Dynamo Models of the Solar Cycle}

\author{Andr\'es Mu\~noz-Jaramillo}
\affil{Department of Physics, Montana State University, Bozeman, MT 59717, USA}
\email{munoz@solar.physics.montana.edu}

\and

\author{Dibyendu Nandy}
\affil{Indian Institute for Science Education and Research, Kolkata, Mohampur 741252, West Bengal, India}
\email{dnandi@iiserkol.ac.in}

\and

\author{Petrus C. H. Martens}
\affil{Department of Physics, Montana State University, Bozeman, MT 59717, USA}
\affil{Harvard-Smithsonian Center for Astrophysics, Cambridge, MA 02138, USA}
\email{martens@solar.physics.montana.edu}

\begin{abstract}

The turbulent magnetic diffusivity in the solar convection zone is one of the most poorly constrained ingredients of mean-field dynamo models. This lack of constraint has previously led to controversy regarding the most appropriate set of parameters, as different assumptions on the value of turbulent diffusivity lead to radically different solar cycle predictions.  Typically, the dynamo community uses double step diffusivity profiles characterized by low values of diffusivity in the bulk of the convection zone.  However, these low diffusivity values are not consistent with theoretical estimates based on mixing-length theory -- which suggest much higher values for turbulent diffusivity.  To make matters worse, kinematic dynamo simulations cannot yield sustainable magnetic cycles using these theoretical estimates.   In this work we show that magnetic cycles become viable if we combine the theoretically estimated diffusivity profile with magnetic quenching of the diffusivity.  Furthermore, we find that the main features of this solution can be reproduced by a dynamo simulation using a prescribed (kinematic) diffusivity profile that is based on the spatiotemporal geometric-average of the dynamically quenched diffusivity.  Here, we provide an analytic fit to the dynamically quenched diffusivity profile, which can be used in kinematic dynamo simulations. Having successfully reconciled the mixing-length theory estimated diffusivity profile with kinematic dynamo models, we argue that they remain a viable tool for understanding the solar magnetic cycle.
\end{abstract}


\section{Introduction}

The solar magnetic cycle involves the recycling of the toroidal and poloidal components of the magnetic field which are generated at spatially segregated source layers that must communicate with each-other (see e.g., Wilmot-Smith et al. 2006\nocite{wilmot-smith-etal06}; Charbonneau 2005\nocite{charbonneau05}). This communication is mediated via magnetic flux-transport, which in most kinematic solar dynamo models, is achieved through diffusive and advective (i.e., by meridional circulation) transport of magnetic fields. The relative strength of turbulent diffusion and meridional circulation determines the regime in which the solar cycle operates, and this has far reaching implications for cycle memory and solar cycle predictions (Yeates, Nandy \& Mackay 2008\nocite{yeates-nandy-mackay08}; Nandy 2010\nocite{nandy10}). As shown in Yeates, Nandy \& Mackay (2008), different assumptions on the strength of turbulent diffusivity in the bulk of the Solar Convection Zone (SCZ) lead to different predictions of the solar cycle (Dikpati, DeToma \& Gilman 2006\nocite{dikpati-detoma-gilman06}; Choudhuri, Chatterjee \& Jiang 2007\nocite{choudhuri-chatterjee-jiang07}). Previously this lack of constraint has led to controversy regarding what value of turbulent diffusivity is more appropriate and yields better solar like solutions (Nandy \& Choudhuri 2002\nocite{nandy-choudhuri02}, Dikpati et al.\ 2002\nocite{dikpati-etal02}, Chatterjee et al.\ 2004\nocite{chatterjee-nandy-choudhuri04}, Dikpati et al.\ 2005\nocite{dikpati-etal05}, Choudhuri et al.\ 2005\nocite{choudhuri-nandy-chatterjee05}). Currently, most dynamo modelers use double-step diffusivity profiles which are somewhat ad-hoc and different from one-another (see Figure~1;
Rempel 2006\nocite{rempel06}, Dikpati and Gilman 2007\nocite{dikpati-gilman07}, Guerrero and de Gouveia Dal Pino 2007\nocite{guerrero-degouveiadalpino07}, Jouve and Brun 2007\nocite{jouve-brun07}).  There is however, a way of theoretically estimating the radial dependence of magnetic diffusivity based on Mixing Length Theory (MLT; Prandtl 1925\nocite{prandtl25}).

\section{Order of Magnitude Estimation}

\begin{figure}
  \begin{center}
  \includegraphics[scale=0.6]{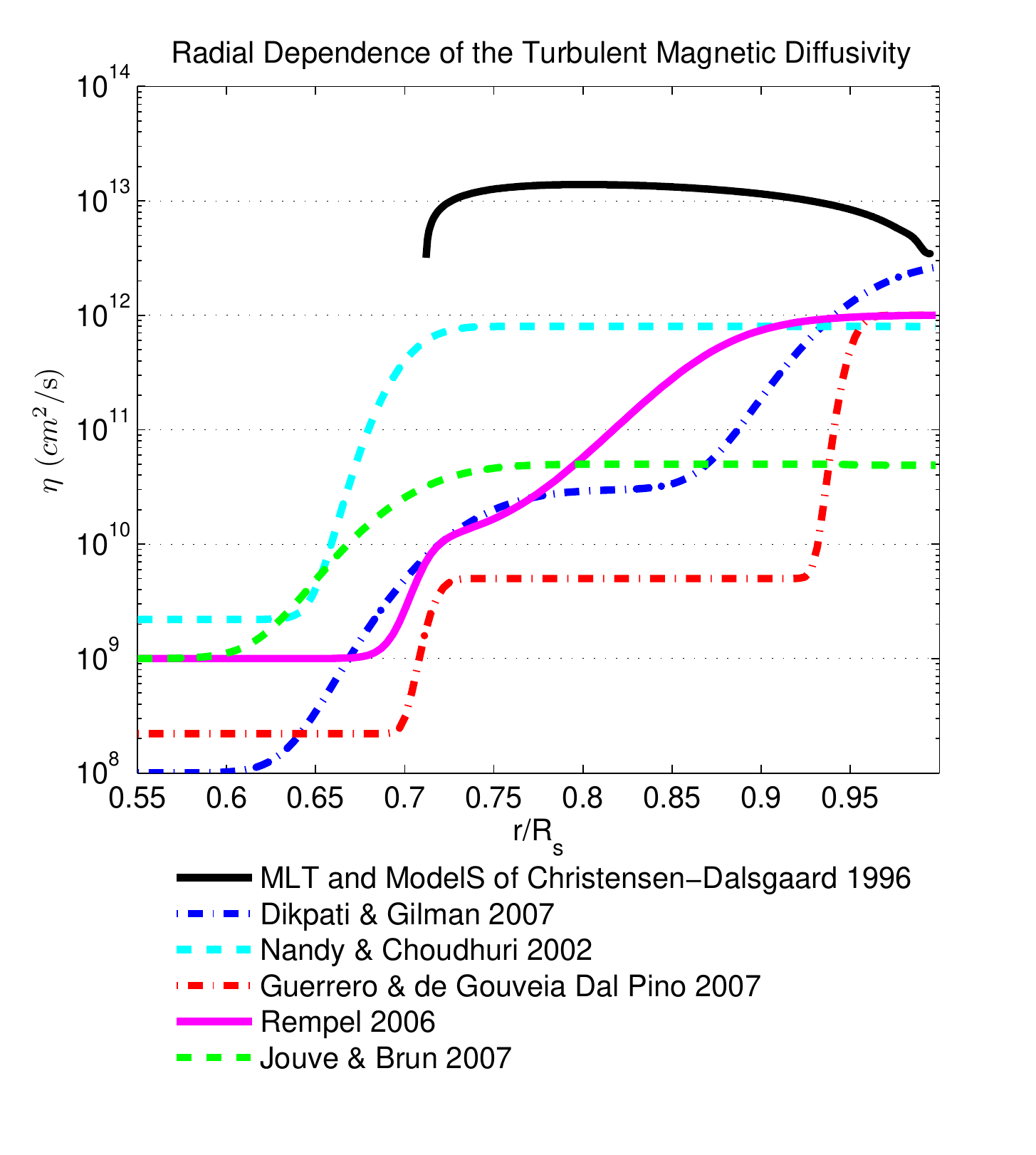}
  \end{center}
  \caption{Different diffusivity profiles used in kinematic dynamo simulations.  The solid black line corresponds to an estimate of turbulent diffusivity obtained by combining Mixing Length Theory (MLT) and the Solar Model S.  The fact that viable solutions can be obtained with such a varied array of profiles have led to debates regarding which profile is more appropriate.  Nevertheless, it is well known that kinematic dynamo simulations cannot yield viable solutions using the MLT estimate.}\label{Fig_1_EtaU}
\end{figure}

Going back to the derivation of the mean-field dynamo equations (after using the first order smoothing approximation), we find that the turbulent diffusivity coefficient becomes (Moffat 1978\nocite{moffat78}):
\begin{equation}\label{EQ_TD}
    \eta = \frac{\tau}{3}\langle v^2 \rangle,
\end{equation}
where $\tau$ is the eddy correlation time and $v$ corresponds to the turbulent velocity field.  In order to make an order of magnitude estimation we turn to MLT, which although not perfect, has been found to be in general agreement with numerical simulations of turbulent convection (Chan \& Sofia 1987\nocite{chan-sofia87}; Abbett et al.\ 1997\nocite{abbett-etal97}).  More specifically we use the Solar Model S (Chistensen-Dalsgaard et al.\ 1996\nocite{Christensen-Dalsgaardetal96}), which is a comprehensive solar interior model used by GONG in all their helioseismic calculations.  Among other quantities, this model estimates the mixing length parameter $\alpha_p$, the convective velocity $v$ for different radii and the necessary variables to calculate the pressure scale height $H_p$.  In terms of those quantities the diffusivity becomes:
\begin{equation}\label{EQ_TD}
    \eta \sim \frac{1}{3}\alpha_p H_p v,
\end{equation}
which we plot in Figure~\ref{Fig_1_EtaU} (solid black line) and show how it compares to commonly used diffusivity profiles.

\section{The Problem and a Possible Solution}

It is evident that there is a major discrepancy between the theoretical estimate and the typical values used inside the convection zone (around two orders of magnitude difference), dynamo models simply cannot operate under such conditions.

A possible solution to this inconsistency resides in the back-reaction that strong magnetic fields have on velocity fields, which results in a suppression of turbulence and thus of turbulent magnetic diffusivity (Roberts \& Soward 1975\nocite{roberts-soward75}).  This magnetic ``quenching" of the turbulent diffusivity has been studied before in different contexts (R\"udiger et al.\ 1994\nocite{ruediger94}; Tobias 1996\nocite{tobias96}; Gilman \& Rempel 2005\nocite{gilman-rempel05}; Mu\~noz-Jaramillo, Nandy \& Martens 2008\nocite{munoz-nandy-martens08}; Guerrero, {Dikpati} \& de Gouveia Dal Pino 2009\nocite{guerrero-etal09}).  However, although this issue has been common knowledge for more than a decade, it's only because of current improvements in computational techniques (Hochbruck \& Lubich 1997\nocite{hochbruck-lubich97}; Hochbruck, Lubich \& Selhofer\nocite{hochbruck-lubich-selhofer98}; Mu\~noz-Jaramillo, Nandy \& Martens 2009; \nocite{munoz-nandy-martens09}MNM09 from here on), that this question can be finally addressed quantitatively.  In this paper we study whether introducing magnetic quenching of the diffusivity can solve this discrepancy and whether the shape of the currently used diffusivity profiles can be understood as a spatiotemporal average of the effective turbulent diffusivity after taking quenching into account.

\section{The Kinematic Mean-Field Dynamo Model}

Our model is based one the axisymmetric dynamo equations:
\begin{eqnarray}\label{25Dynamo}
  \frac{\partial A}{\partial t} + \frac{1}{s}\left[ \textbf{v}_p \cdot \nabla (sA) \right] &=& \eta\left( \nabla^2 - \frac{1}{s^2}  \right)A \nonumber\\
  & & \\
  \frac{\partial B}{\partial t}  + s\left[ \textbf{v}_p \cdot \nabla\left(\frac{B}{s} \right) \right] + (\nabla \cdot \textbf{v}_p)B&=& \eta\left( \nabla^2 - \frac{1}{s^2}  \right)B + s\left(\left[ \nabla \times (A\bf \hat{e}_\phi) \right]\cdot \nabla \Omega\right)  \nonumber \\
  & & + \frac{1}{s}\frac{\partial \eta}{\partial r}\frac{\partial (sB)}{\partial r}
  + \frac{1}{s r^2}\frac{\partial \eta}{\partial \theta}\frac{\partial (s B)}{\partial \theta}, \nonumber
\end{eqnarray}
where A is the $\phi$-component of the potential vector (from which $B_r$ and $B_\theta$ can be obtained), B is the toroidal field ($B_\phi$), $v_p$ is the meridional flow, $\Omega$ the differential rotation, $\eta$ the turbulent magnetic diffusivity and $s = r\sin(\theta)$.  In order to integrate these equations we need to prescribe four ingredients: meridional flow, differential rotation, poloidal field regeneration mechanism and turbulent magnetic diffusivity.  We use the same meridional flow profile we defined in MNM09, which better captures the features present in helioseismic data.  Our flow profile has a penetration depth of $.675R_\odot$ and a top speed of $12m/s$.  For the differential rotation we use the analytical form of Charbonneau et al.\ (1999\nocite{charbonneau-etal99}), with a tachocline centered at $0.7R_\odot$ and whose thickness is $0.05R_\odot$.  For the poloidal field regeneration mechanism we use the improved ring-duplet algorithm described in Nandy, Mu\~noz-Jarmillo \& Martens 2010\nocite{nandy-munoz-martens10} (NMM10) and Mu\~noz-Jaramillo et al. 2010\nocite{munoz-etal10} (MNMY10), but using a value of $K_0 = 3900$, in order to insure super-criticality.  As in NMM10 and MNMY10, we restrict active region emergence to latitudes between $45^oN$ and $45^oS$. Specifics about our treatment of the turbulent diffusivity are described below.  More details regarding kinematic dynamo models can be found in a review by Charbonneau (2005\nocite{charbonneau05}) and references therein.

\subsection{Turbulent Magnetic Diffusivity and Diffusivity Quenching}

In order to study the effect of magnetic quenching on dynamo models we introduce an additional state variable $\eta_{mq}$ governed by the following differential equation:
\begin{equation}\label{Eta_st}
    \frac{\partial\eta_{mq}}{\partial t} = \frac{1}{\tau}\left(\frac{\eta_{MLT}(r)}{1+\textbf{B}^2(r,\theta,t)/B_0^2} - \eta_{mq}(r,\theta,t) \right).
\end{equation}
In a steady state, $\eta_{mq}$ corresponds to the MLT estimated diffusivity $\eta_{MLT}(r)$ quenched in such a way that the diffusivity is halved for a magnetic field of amplitude $B_0=6700$ Gauss (G).  This value corresponds to the average equipartition field strength inside the SCZ calculated using the Solar Model S. The characteristic time of relaxation $\tau = 30$ days is an estimate of the average eddy turnover time.

\begin{figure}
  \begin{center}
  \includegraphics[scale=0.6]{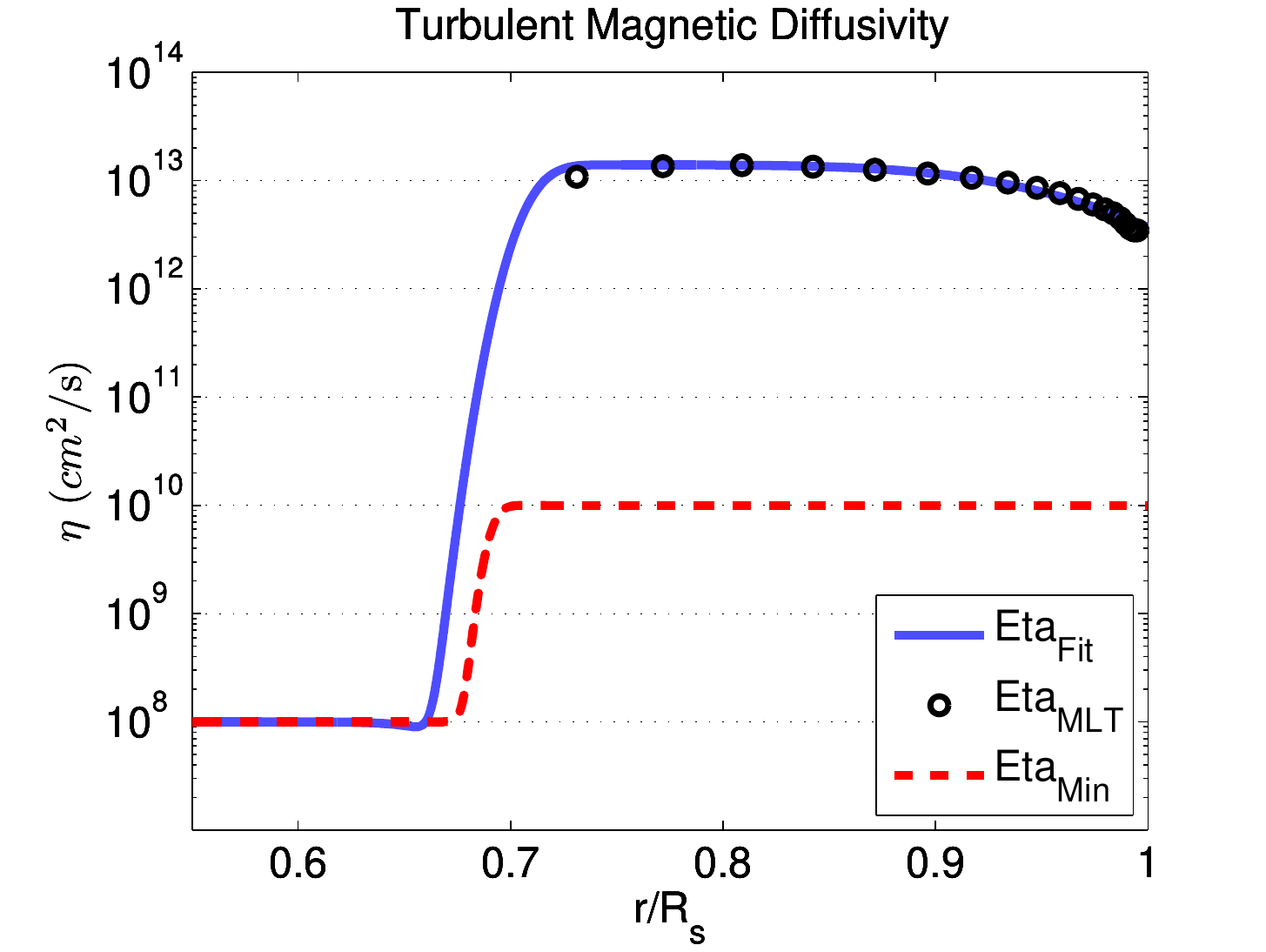}
  \end{center}
\caption{Fit (solid line) of diffusivity as a function of radius to the mixing-length theory estimate (circles).  As part of our definition of effective diffusivity we put a limit on how much can the diffusivity be quenched.  This minimum diffusivity has a radial dependence that is shown as a dashed line.}\label{Fig_2_EtaC}
\end{figure}

We make a fit of $\eta_{MLT}(r)$ using the following analytical profile (see Figure~\ref{Fig_2_EtaC}):
\begin{equation}\label{Eq_Eta_MLT}
      \eta(r) = \eta_0 + \frac{\eta_1 - \eta_0}{2}\left( 1 + \operatorname{erf}\left( \frac{r - r_1}{d_1}  \right)
      \right) + \frac{\eta_2 - \eta_1 - \eta_0}{2}\left( 1 + \operatorname{erf}\left( \frac{r - r_2}{d_2},  \right) \right)
\end{equation}
where $\eta_0 = 10^8 cm^2/s$ corresponds to the diffusivity at the bottom of the
computational domain; $\eta_1 = 1.4\times10^{13} cm^2/s$ and $\eta_2 = 10^{10} cm^2/s$ control the diffusivity in the
convection zone; $r_1 = 0.71R_\odot$, $d_1 = 0.015R_\odot$, $r_2 = 0.96R_\odot$ and $d_2 = 0.09R_\odot$ characterize the transitions from one value of diffusivity to the other.  With this in mind, we define the effective diffusivity at any given point as
\begin{equation}\label{Eta_eff}
    \eta_{eff}(r,\theta,t) = \eta_{min}(r) + \eta_{mq}(r,\theta,t).
\end{equation}
with the minimum magnetic diffusivity $\eta_{min}(r)$ given by the following analytical profile (see Figure~\ref{Fig_2_EtaC}):
\begin{equation}\label{Eta_Min}
      \eta_{min}(r) = \eta_0 + \frac{\eta_{cz} - \eta_0}{2}\left( 1 + \operatorname{erf}\left( \frac{r - r_{cz}}{d_{cz}}  \right)\right),
\end{equation}
where $\eta_{cz} = 10^{10} cm^2/s$, $r_{cz} = 0.69R_\odot$, and $d_{cz} = 0.07R_\odot$.  Since diffusivity is now a state variable, small errors can lead to negative values of diffusivity, which in turn leads to unbound magnetic field growth.  By putting a limit on how small can the diffusivity become, we successfully avoid this type of computational instability.

\section{Dynamo Simulations}

We use the SD-Exp4 code (see the Appendix of MNM09) to solve the dynamo equations (Eqs. \ref{25Dynamo}).  Our computational domain comprises the SCZ   and upper layer of the solar radiative zone in the northern hemisphere ($0.55R_\odot\leq r\leq R_\odot$ and $0\leq\theta\leq\pi$).  In order to approximate the spatial differential operators with finite differences we use a uniform grid (in radius and co-latitude), with a resolution of $400\times400$ gridpoints.

Our boundary conditions assume that the magnetic field is anti-symmetric across the equator ($\partial A/\partial\theta|_{\theta = \pi/2} = 0$; $\partial B/\partial\theta|_{\theta = \pi/2} = 0$), that the plasma below the lower boundary is a perfect conductor ($A(r = 0.55R_\odot,\theta) = 0$; $\partial (r B)/\partial r|_{r = 0.55R_\odot} = 0$), that the magnetic field is axisymmetric ($A(r,\theta=0)=0$; $B(r,\theta=0)=0$), and that field at the surface is radial ($\partial (r A)/\partial r|_{r = R_\odot} = 0$; $B(r = R_\odot,\theta) = 0$).  Our initial conditions consist of a large toroidal belt and no poloidal component.  After setting up the problem we let the magnetic field evolve for 200 years allowing the dynamo to reach a stable cycle.

\section{Results and Discussion}

The first important result is the existence of a uniform cycle in dynamic equilibrium.  The presence of a diffusivity quenching algorithm allows the dynamo to become viable in a regime in which kinematic dynamo models cannot operate thanks to the creation of pockets of relatively low magnetic diffusivity (where long lived magnetic structures can exist).  This can be clearly seen in Figure~\ref{Fig_3_MEvol}, which shows snapshots of the effective turbulent diffusivity and the toroidal and poloidal components of the magnetic field at different moments during the sunspot cycle (half a magnetic cycle).  As expected the turbulent diffusivity is strongly suppressed by the magnetic field (especially by the toroidal component), increasing the diffusive timescale to the point where diffusion and advection become equally important for flux transport dynamics.  This slow-down of the diffusive process is crucial for the survival of the magnetic cycle since it gives differential rotation more time to amplify the weak poloidal components of the magnetic field into strong toroidal belts, while providing them a measure of isolation from the top ($r = R_\odot$) and polar ($\theta=0$) boundary conditions (B=0).

\subsection{Time and Spatiotemporal Averages of the Effective Diffusivity}

\begin{figure}[c]
  \begin{tabular}{ccc}
  Effective Diffusivity                 & Toroidal Field                        & Poloidal Field \\
  \includegraphics[scale=0.35]{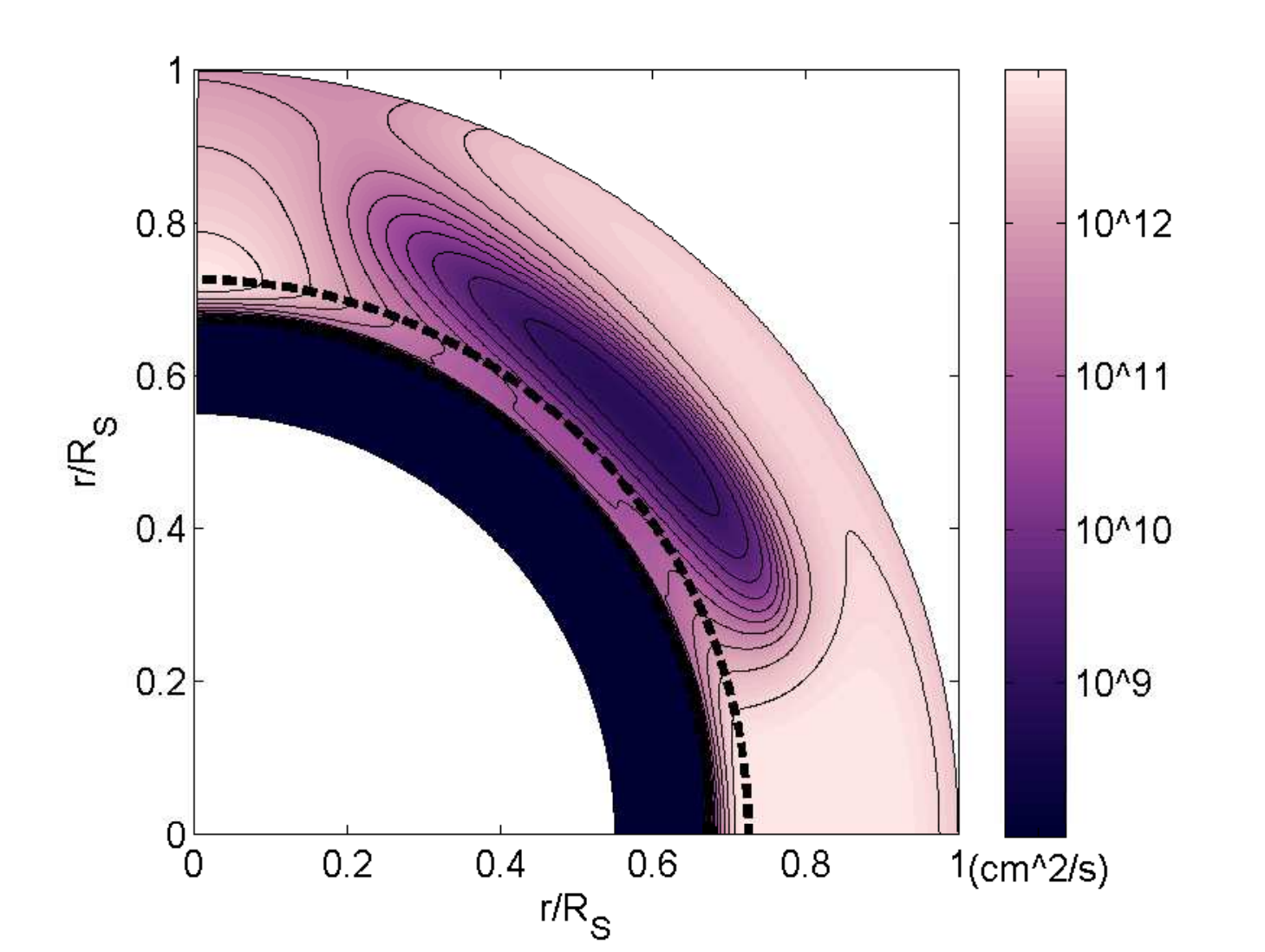} & \includegraphics[scale=0.35]{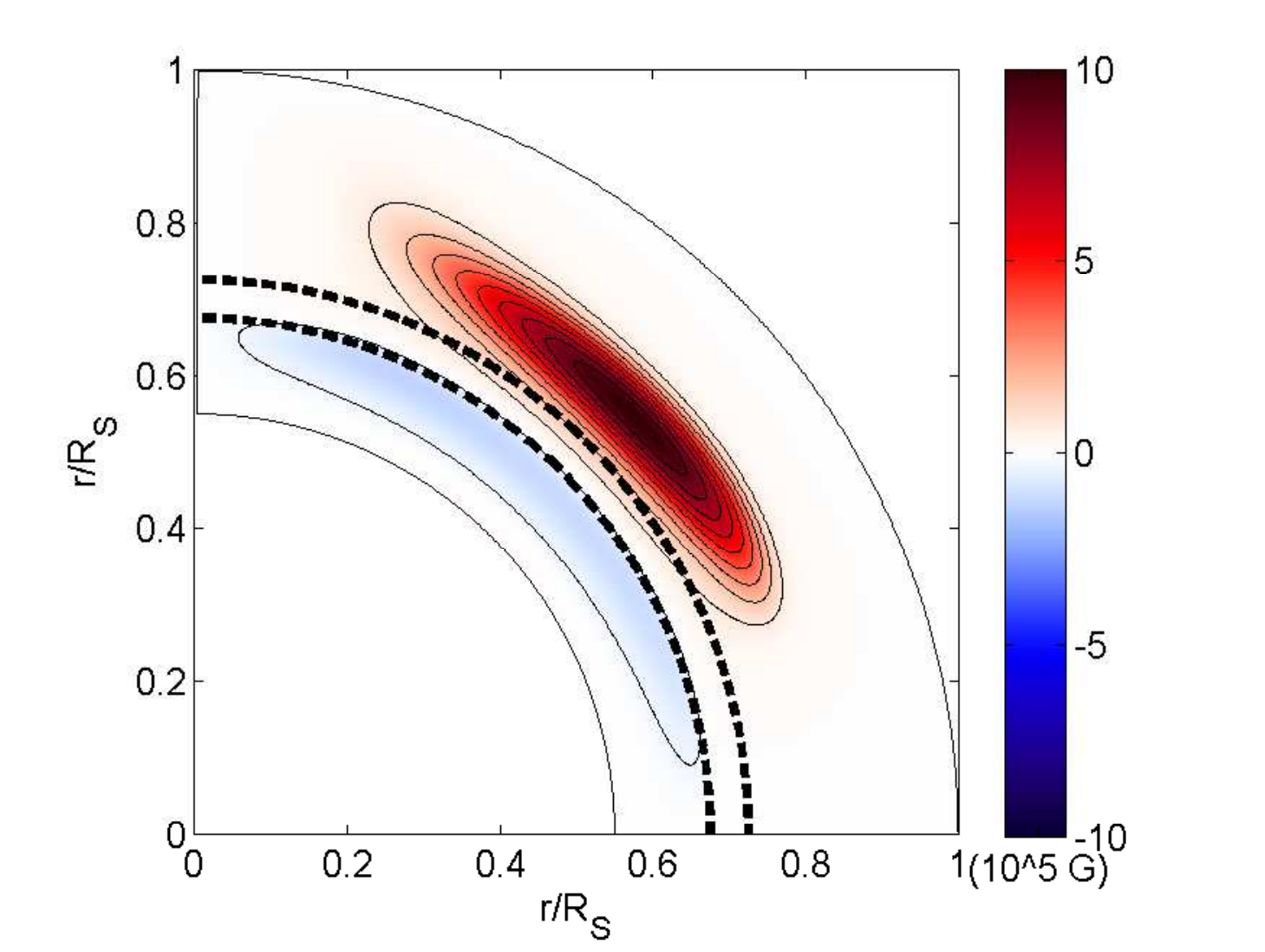} & \includegraphics[scale=0.35]{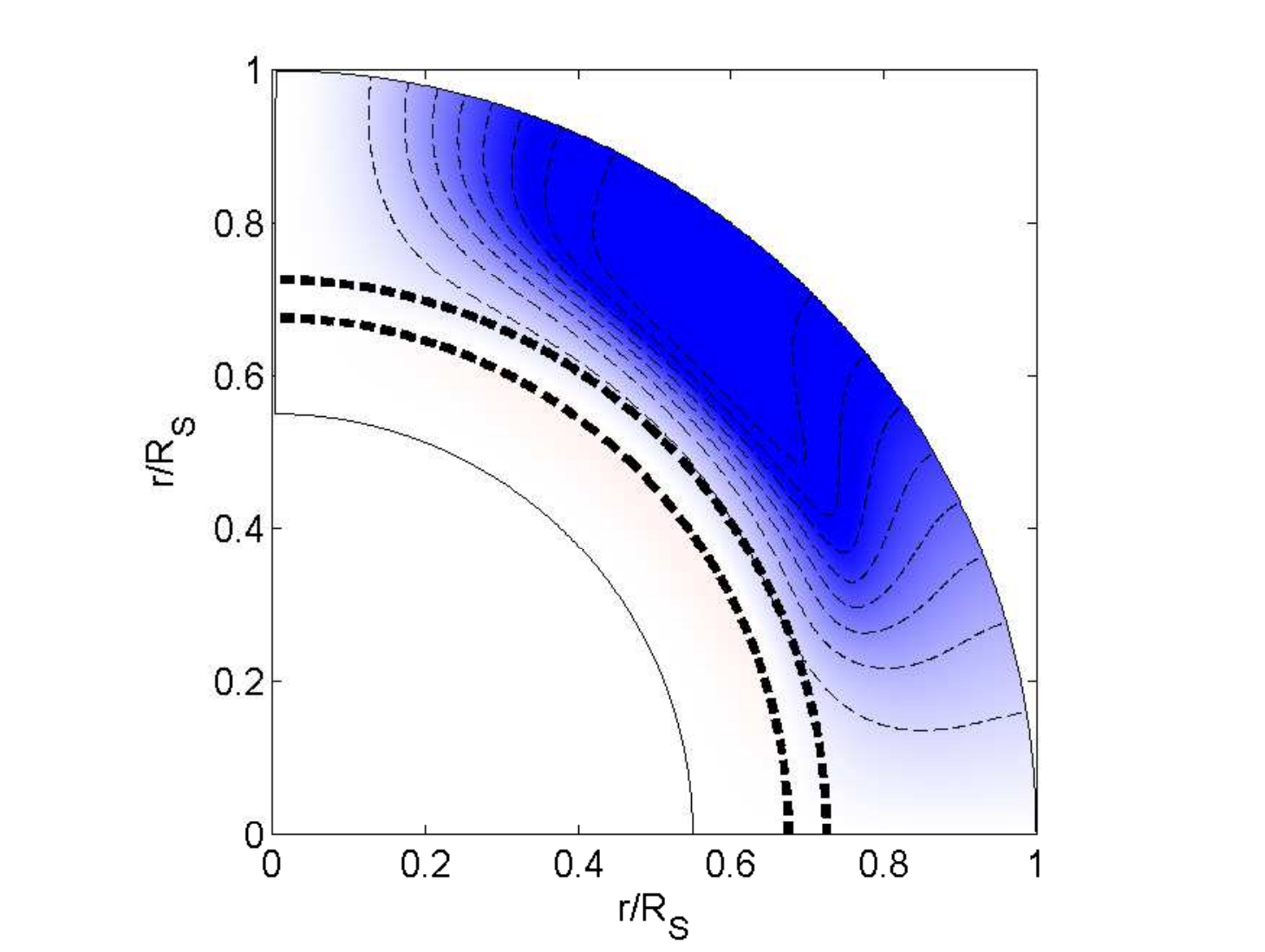} \\
  \includegraphics[scale=0.35]{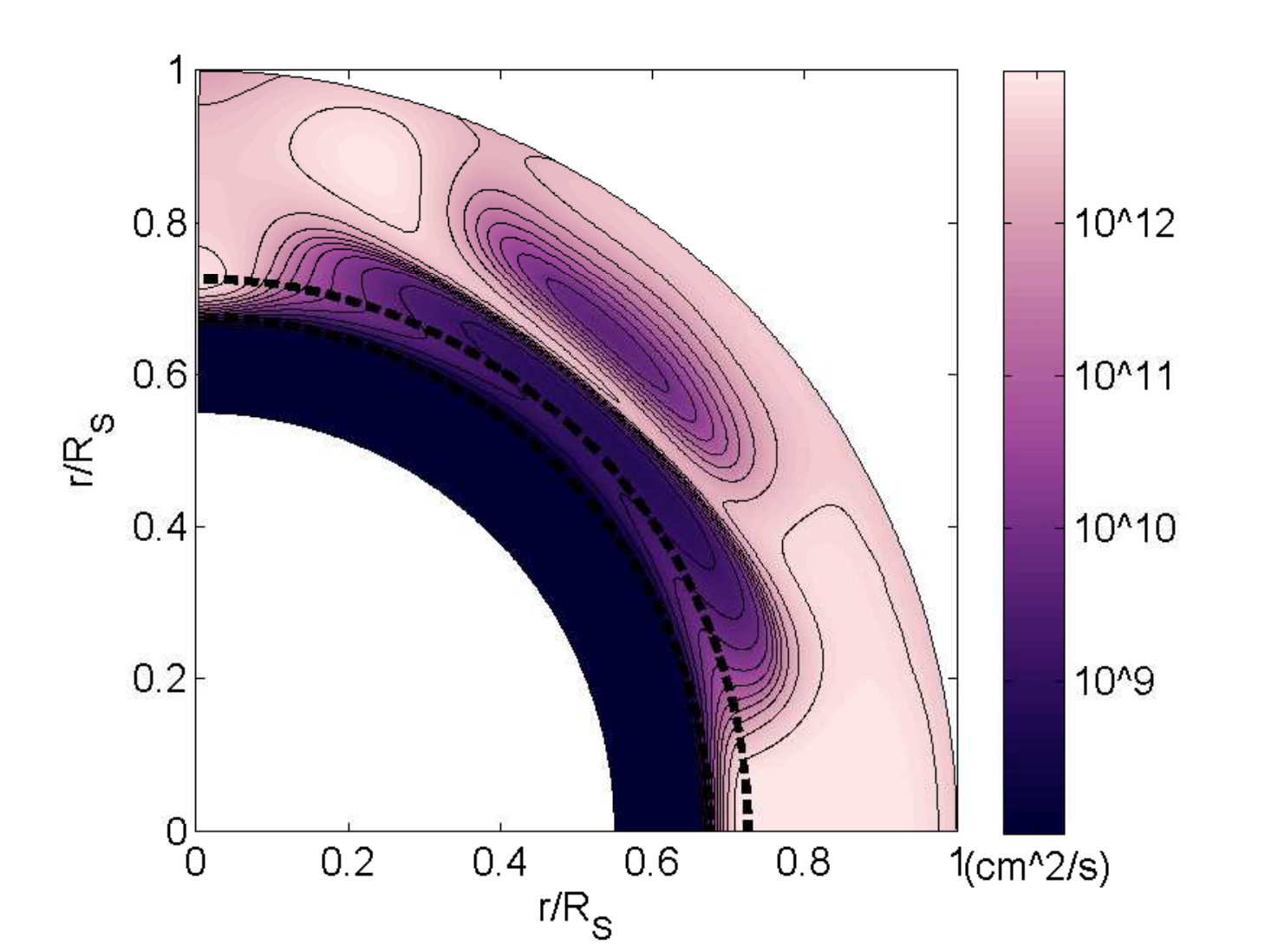} & \includegraphics[scale=0.35]{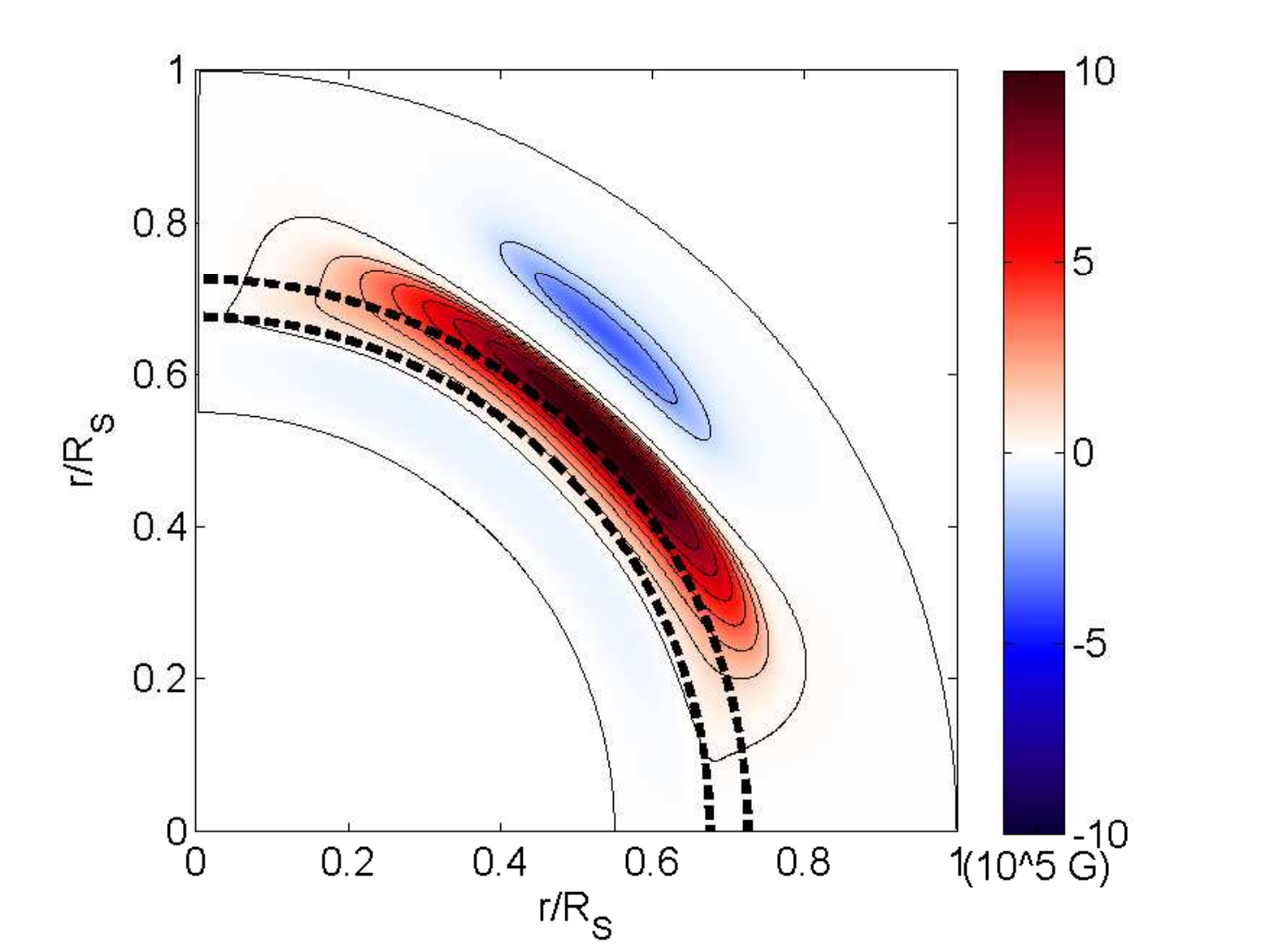} & \includegraphics[scale=0.35]{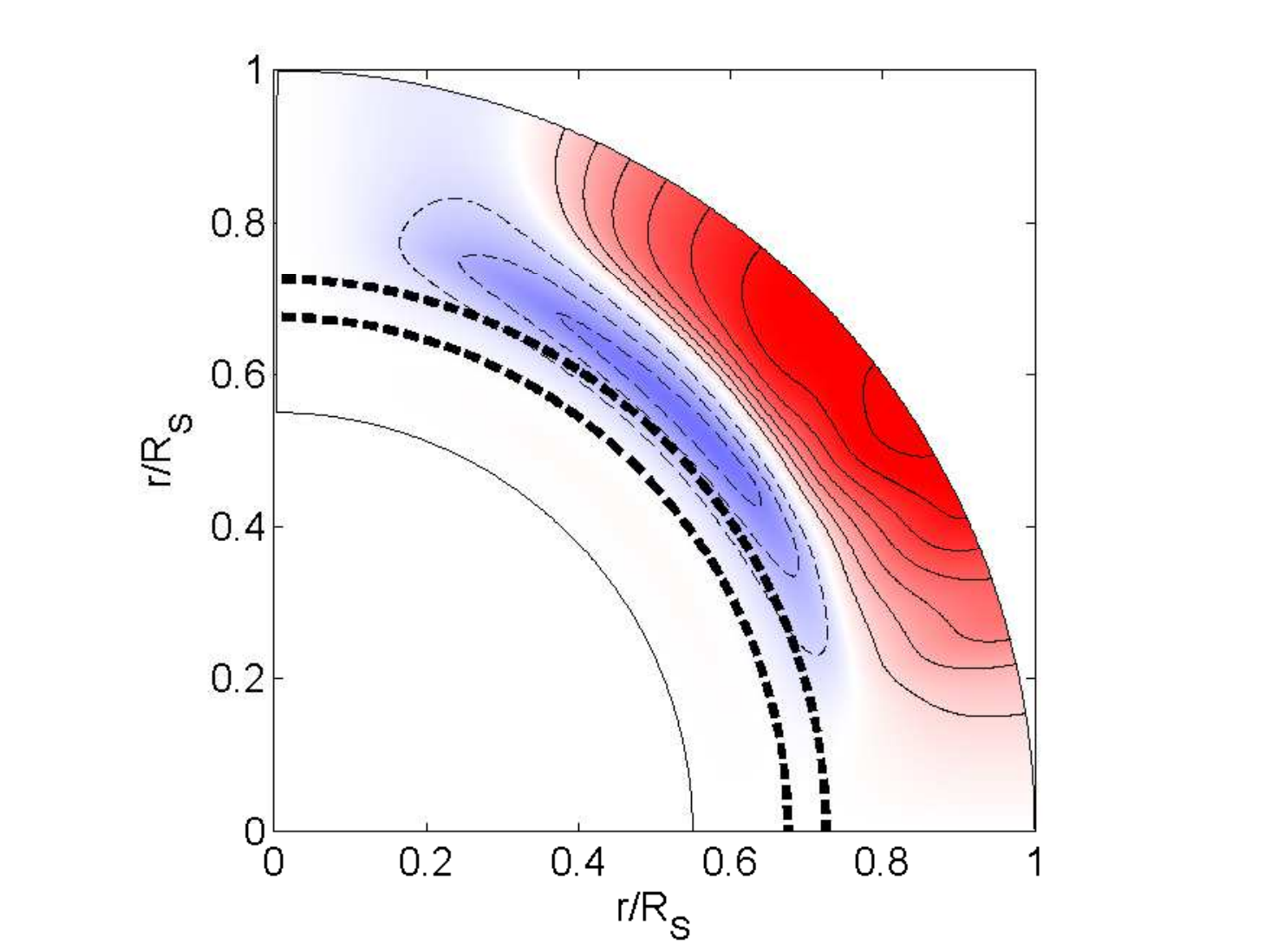} \\
  \includegraphics[scale=0.35]{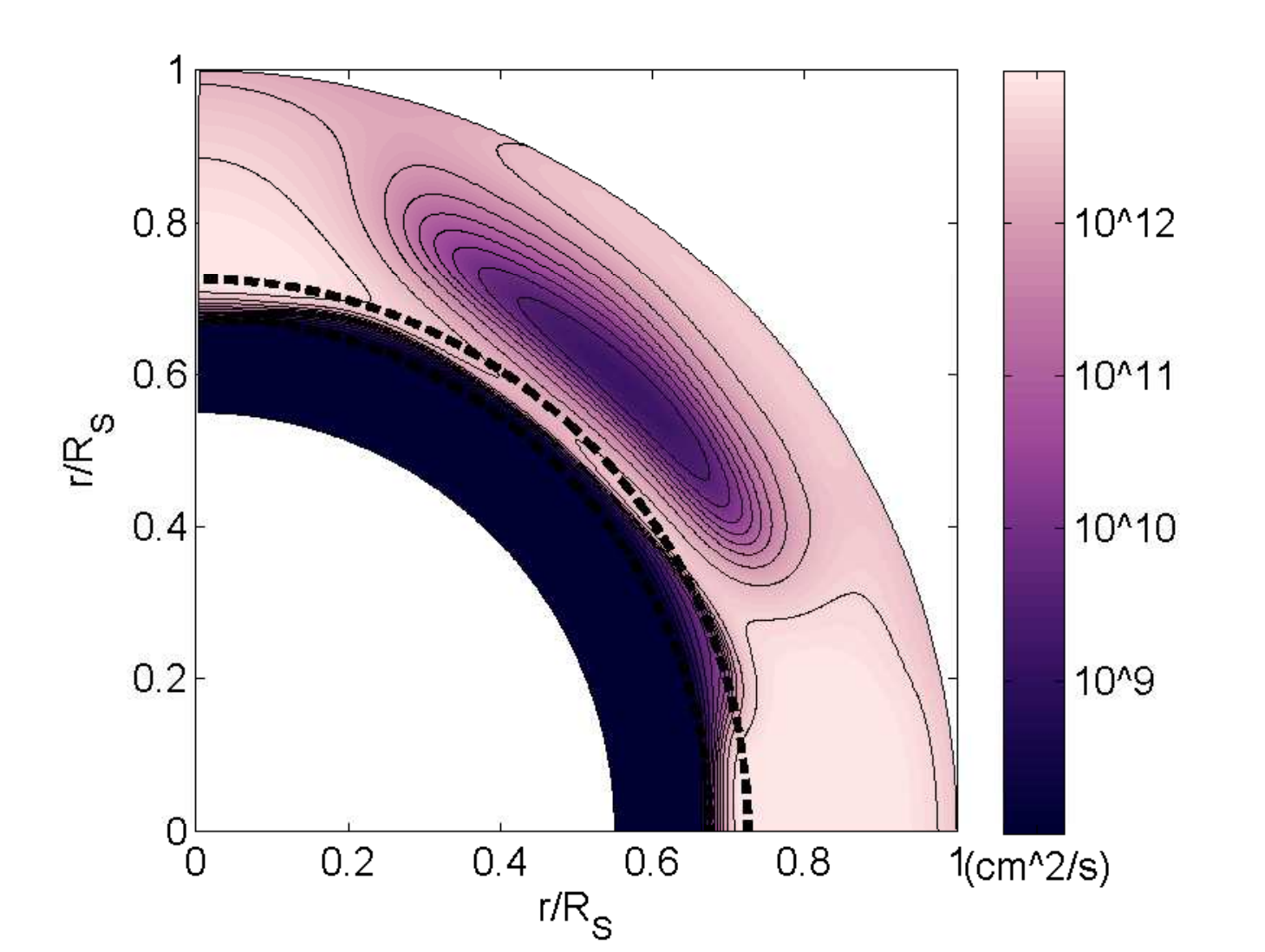} & \includegraphics[scale=0.35]{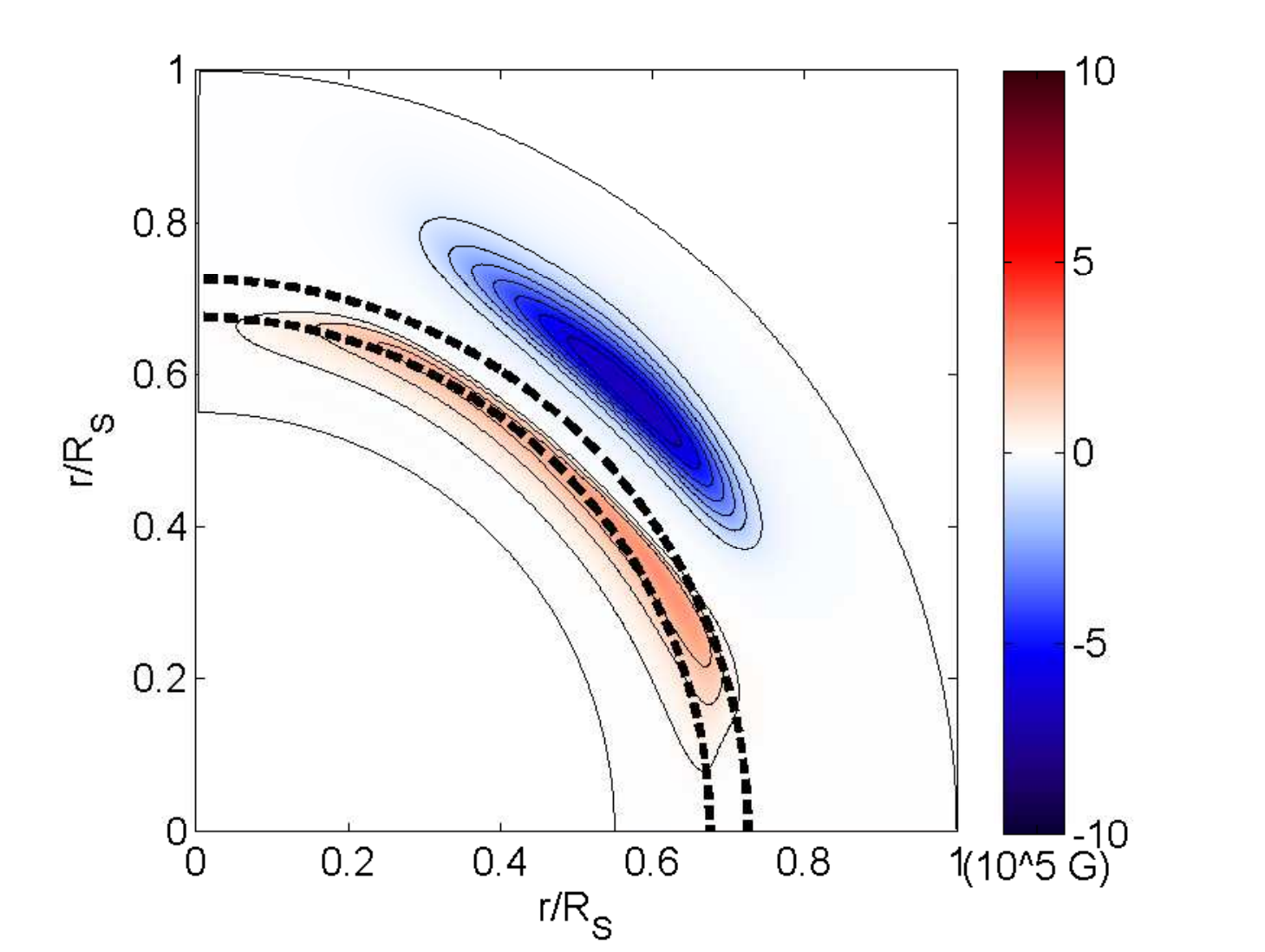} & \includegraphics[scale=0.35]{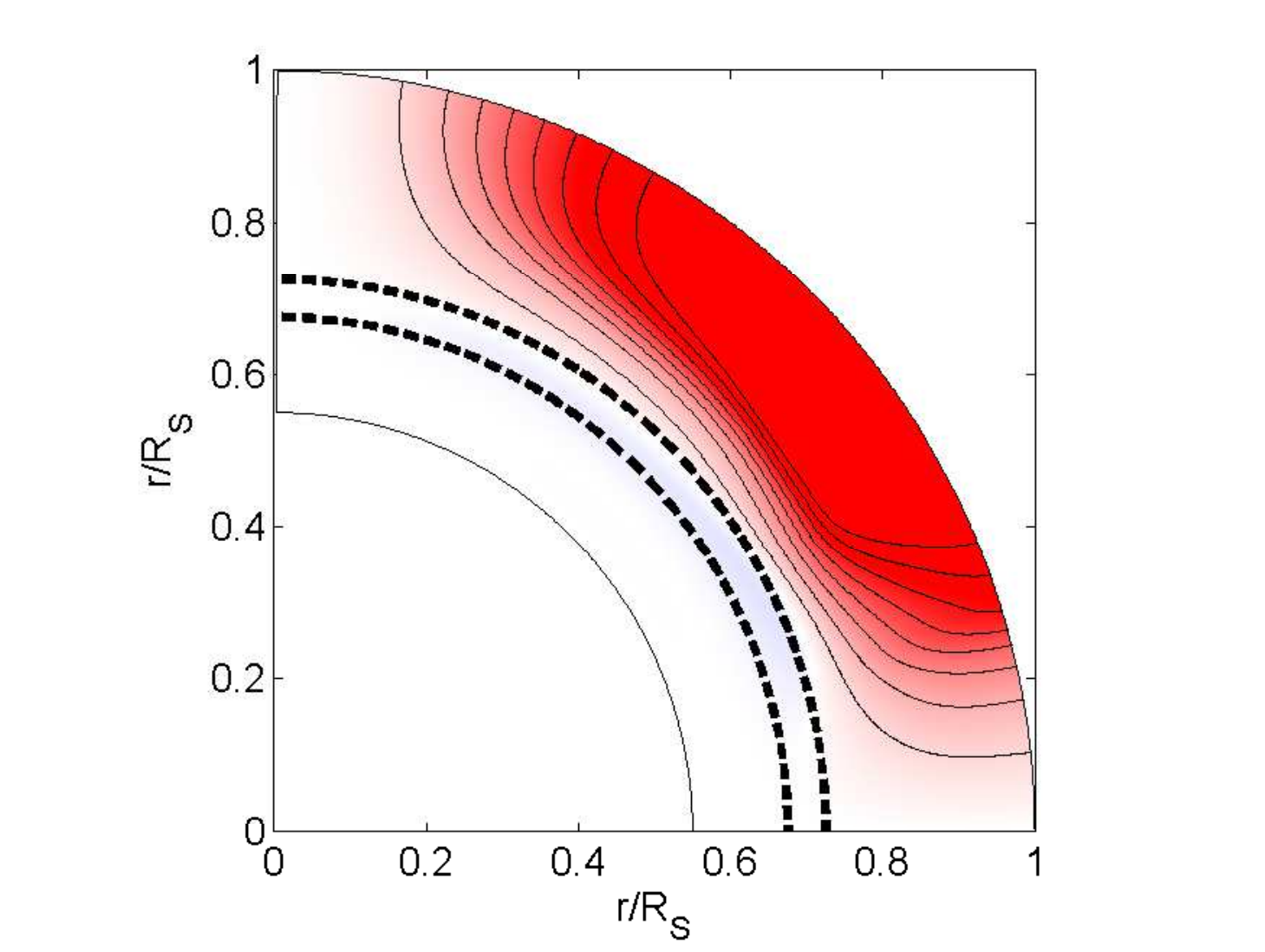}
  \end{tabular}
\caption{Snapshots of the effective diffusivity and the magnetic field over half a dynamo cycle (a
sunspot cycle).  For the poloidal field a solid (dashed) line corresponds to clockwise (counter-clockwise) poloidal field lines. Each row is advanced in time by an sixth of the dynamo cycle (a third of the sunspot
cycle) i.e., from top to bottom $t = 0, \tau/6, \tau/3$ and $\tau/2$. As expected, the turbulent diffusivity is strongly depressed by the magnetic field (especially by the toroidal component).  This reduces the diffusive time-scale to a point where the magnetic cycle becomes viable and sustainable.}\label{Fig_3_MEvol}
\end{figure}

\begin{figure}[c]
\begin{tabular}{c}
  \begin{tabular}{cc}
  Geometric Time Average of $\eta_{eff}$ & Arithmetic Time Average of $\eta_{eff}$\\
  \includegraphics[scale=0.5]{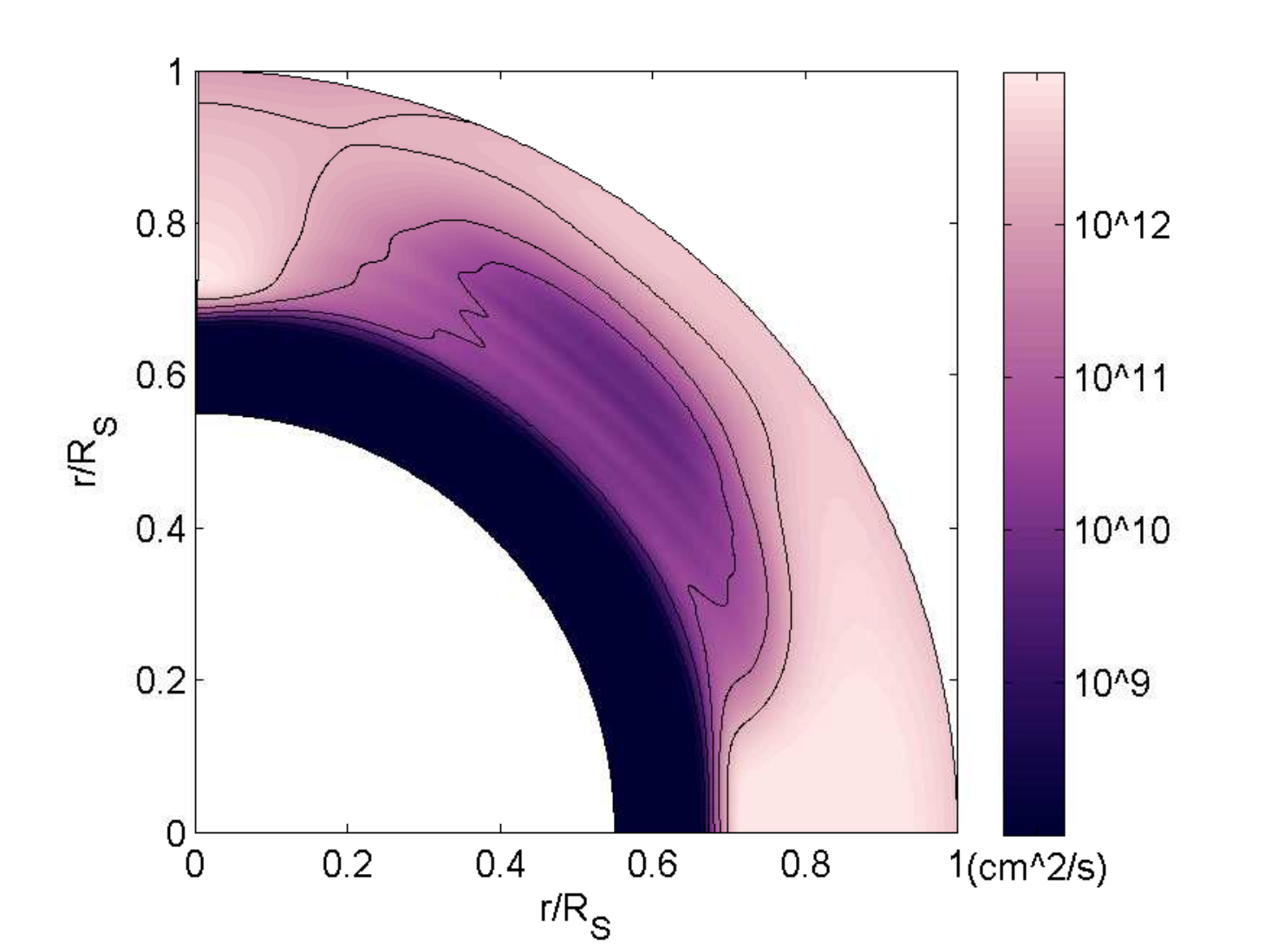} & \includegraphics[scale=0.5]{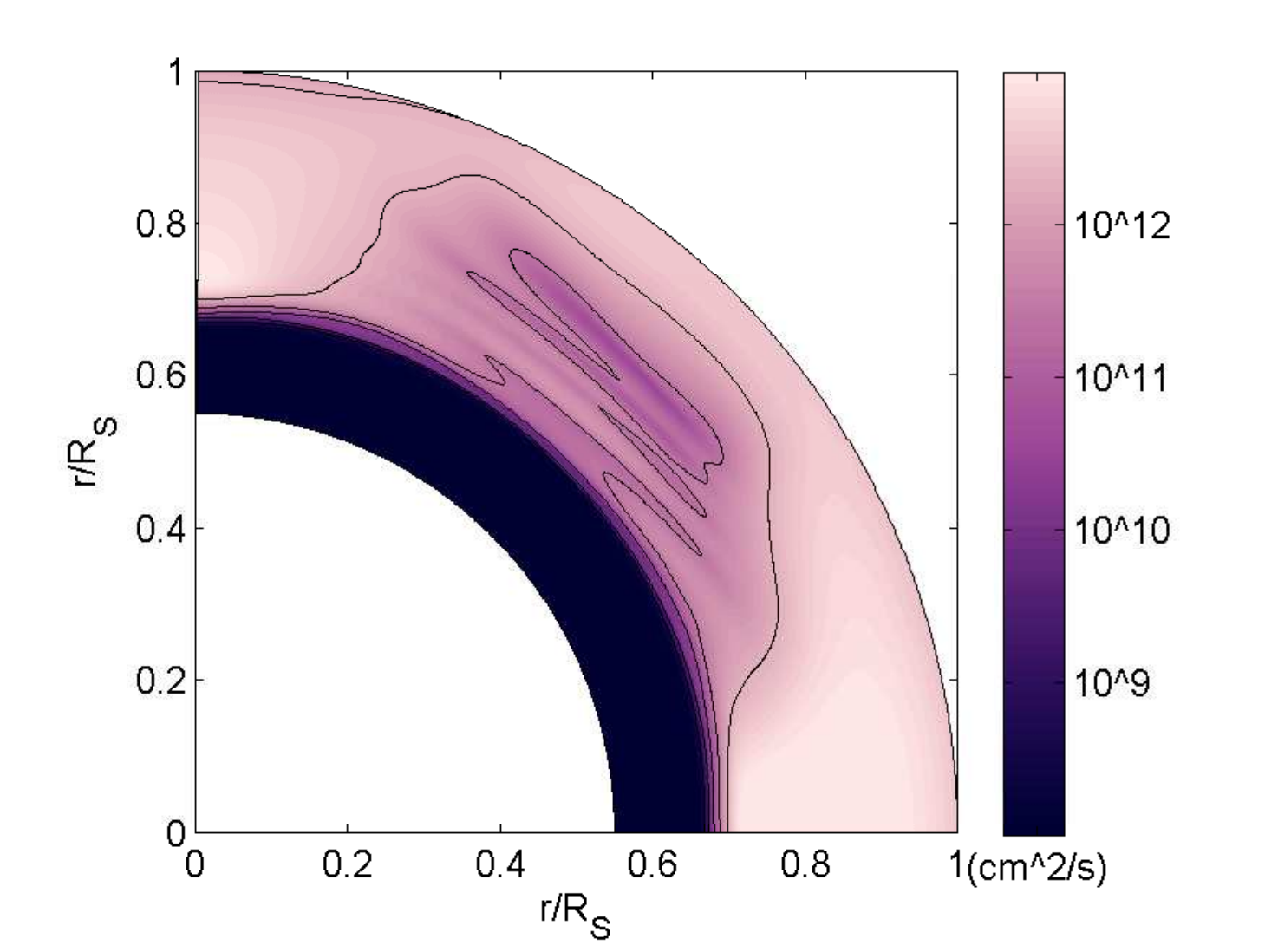} \\
                (a)                    &                     (b)
  \end{tabular}\\
  Spatiotemporal Averages of $\eta_{eff}$\\
  \includegraphics[scale=0.5]{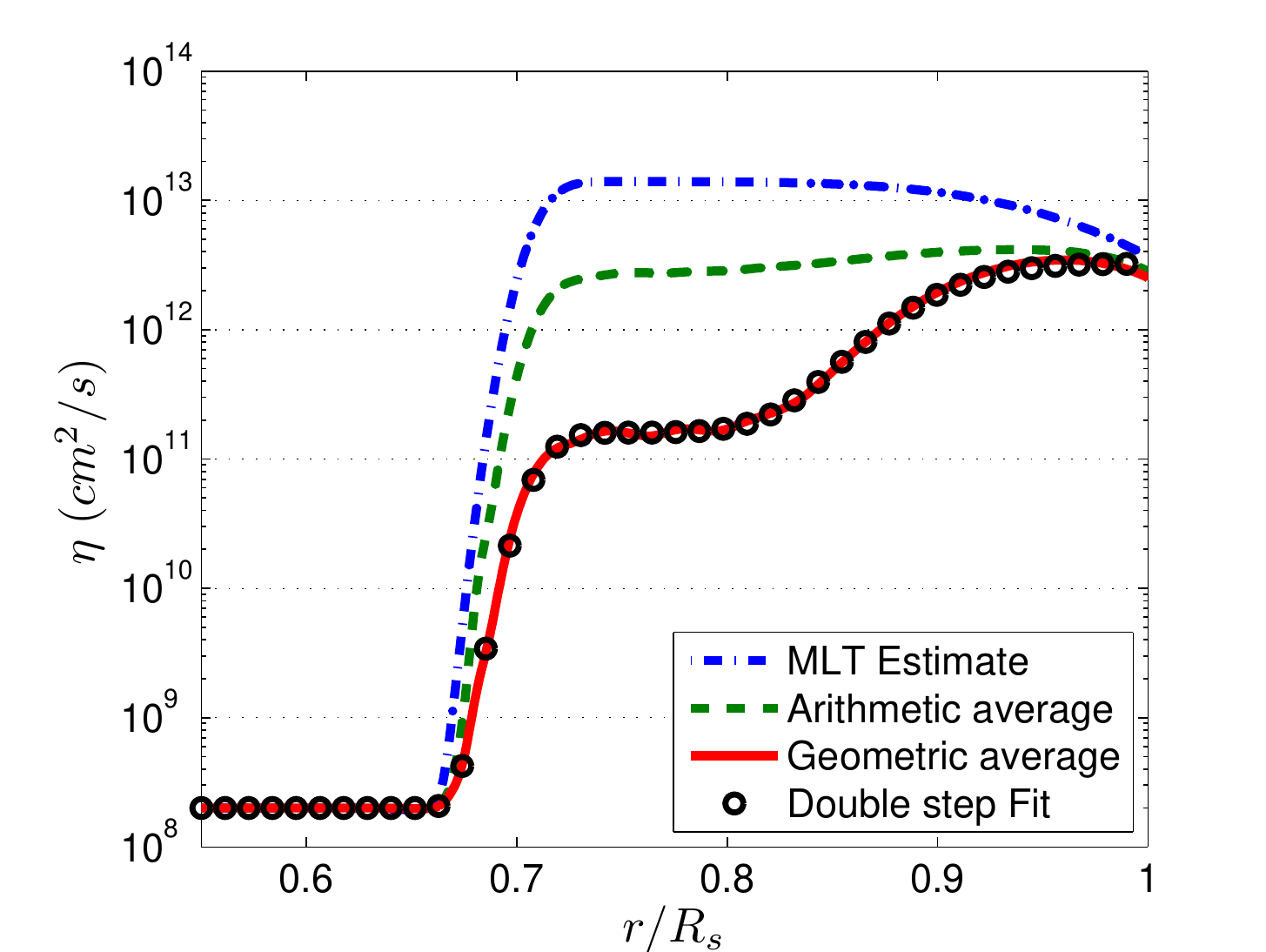}\\
  (c)
\end{tabular}
\caption{Geometric (Eq.~\ref{Eq_ArmAv}) and arithmetic (Eq.~\ref{Eq_GeoAv}) averages of the effective diffusivity.  We find that the geometric time average (a) and geometric spatiotemporal average (solid line in c) capture the essence of diffusivity quenching much more accurately than the arithmetic time average (b) and arithmetic spatiotemporal average (dashed line in c).  We then use a fit to the geometric spatiotemporal average (circles in c), in a kinematic dynamo simulation in order to find out how it compares with the simulation that uses the MLT estimate and magnetic quenching.}\label{Fig_4_Avr}
\end{figure}

Given that ultimately we want to understand how adequate kinematic diffusivity profiles are and whether they are plausible representations of physical reality consistent with MLT, we need to find a connection between kinematic profiles and the dynamically quenched diffusivity.  Because of this, the next natural step is to find time and spatiotemporal averages of the effective diffusivity.  After careful consideration we have found that the geometric average (also known as logarithmic average):
\begin{equation}\label{Eq_GeoAv}
    \begin{array}{c}
      \log[\eta_{avT}(r_i,\theta_j)] = \frac{1}{N_t}\displaystyle\sum_{n=0}^{N_t} \log[\eta_{eff}(r_i,\theta_j,t_n)]\\
       \\
      \log[\eta_{avTL}(r_i)] = \frac{1}{N_\theta}\displaystyle\sum_{j=0}^{N_\theta} \log[\eta_{avT}(r_i,\theta_j)].
    \end{array}
\end{equation}
shown in Figure~\ref{Fig_4_Avr}-a (time average) and as a solid line in Figure~\ref{Fig_4_Avr}-c (spatiotemporal average), captures better the physical and mathematical nuances of diffusive processes than the arithmetic average:
\begin{equation}\label{Eq_ArmAv}
    \begin{array}{c}
      \eta_{avT}(r_i,\theta_j) = \frac{1}{N_t}\displaystyle\sum_{n=0}^{N_t} \eta_{eff}(r_i,\theta_j,t_n)\\
       \\
       \eta_{avTL}(r_i) = \frac{1}{N_\theta}\displaystyle\sum_{j=0}^{N_\theta} \eta_{avT}(r_i,\theta_j),
    \end{array}
\end{equation}
shown in Figure~\ref{Fig_4_Avr}-b (time average) and as a solid line in Figure~\ref{Fig_4_Avr}-c (spatiotemporal average).  The reason is that the geometric average is more appropriate for systems whose evolution is exponential in nature and when the values involved in the average are dependent of each other.  Since the diffusion equation has exponential solutions and our intention is to capture the essence of a cyclic behavior in which the magnetic field is diffused and advected on a closed circuit (thus there is a direct dependence between the different elements considered in our average), the geometric average is the best choice.  As an additional advantage, the geometric mean gives more weight to small values than the arithmetic mean; this is important because by definition the regions of depressed diffusivity are also those which the magnetic field visits during the cycle.  Figure~\ref{Fig_3_MEvol} shows that turbulent diffusivity is being suppressed mainly in a region centered at mid-latitudes, whereas the polar and equatorial regions remain unquenched (and devoid of magnetic field).  If one then compares the geometric time average (Fig.~\ref{Fig_4_Avr}-a) and the arithmetic time average (Fig.~\ref{Fig_4_Avr}-a) with the evolution of the magnetic diffusivity (Fig.~\ref{Fig_3_MEvol}), it's clear that the geometric average is qualitatively truer to the essence of the process.  As we will see below, the quantitative results are very encouraging as well.

\subsection{Comparison with Kinematic Dynamo Simulations}

Once we calculate the time and latitude geometric average of the effective diffusivity we obtain a radial diffusivity profile (see solid line on Fig.~\ref{Fig_4_Avr}-c), which interestingly can be accurately described as a double-step profile (Eq.~\ref{Eq_Eta_MLT}) with the following parameters: $\eta_0 = 10^8 cm^2/s$, $\eta_1 = 1.6\times10^{11} cm^2/s$, $\eta_2 = 3.25\times10^{12} cm^2/s$, $r_1 = 0.71R_\odot$, $d_1 = 0.017R_\odot$, $r_2 = 0.895R_\odot$ and $d_2 = 0.051R_\odot$ (see circles on Fig.~\ref{Fig_4_Avr}-c).  We then run a kinematic dynamo simulation (no magnetic quenching), leaving all ingredients intact with the exception of the turbulent diffusivity profile -- for which we use this double-step fit.

\begin{figure}[c]
\begin{center}
  \begin{tabular}{c}
  Surface Radial Field and AR emergence\\
  \includegraphics[scale=0.5]{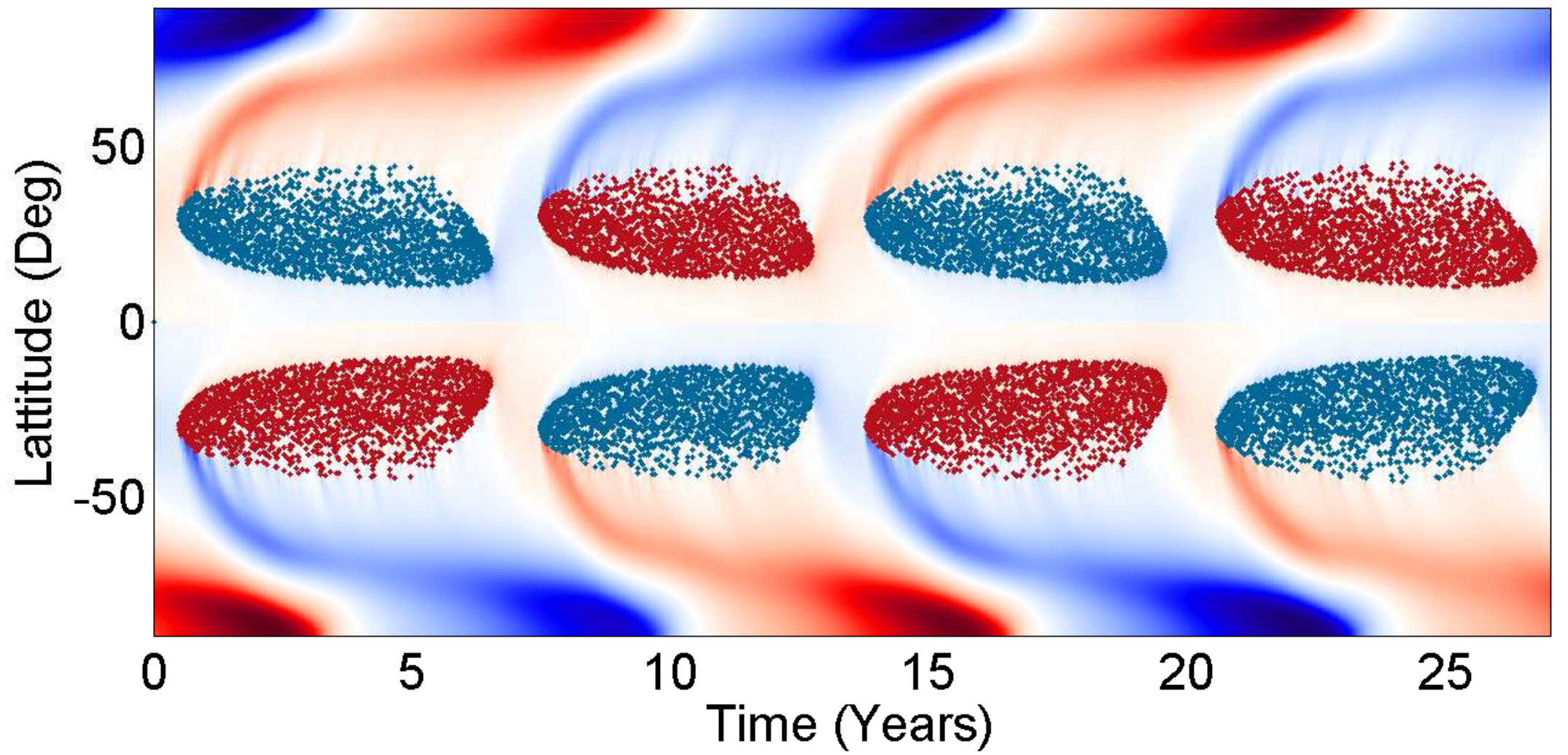}\\
                 (a)                  \\
  \includegraphics[scale=0.5]{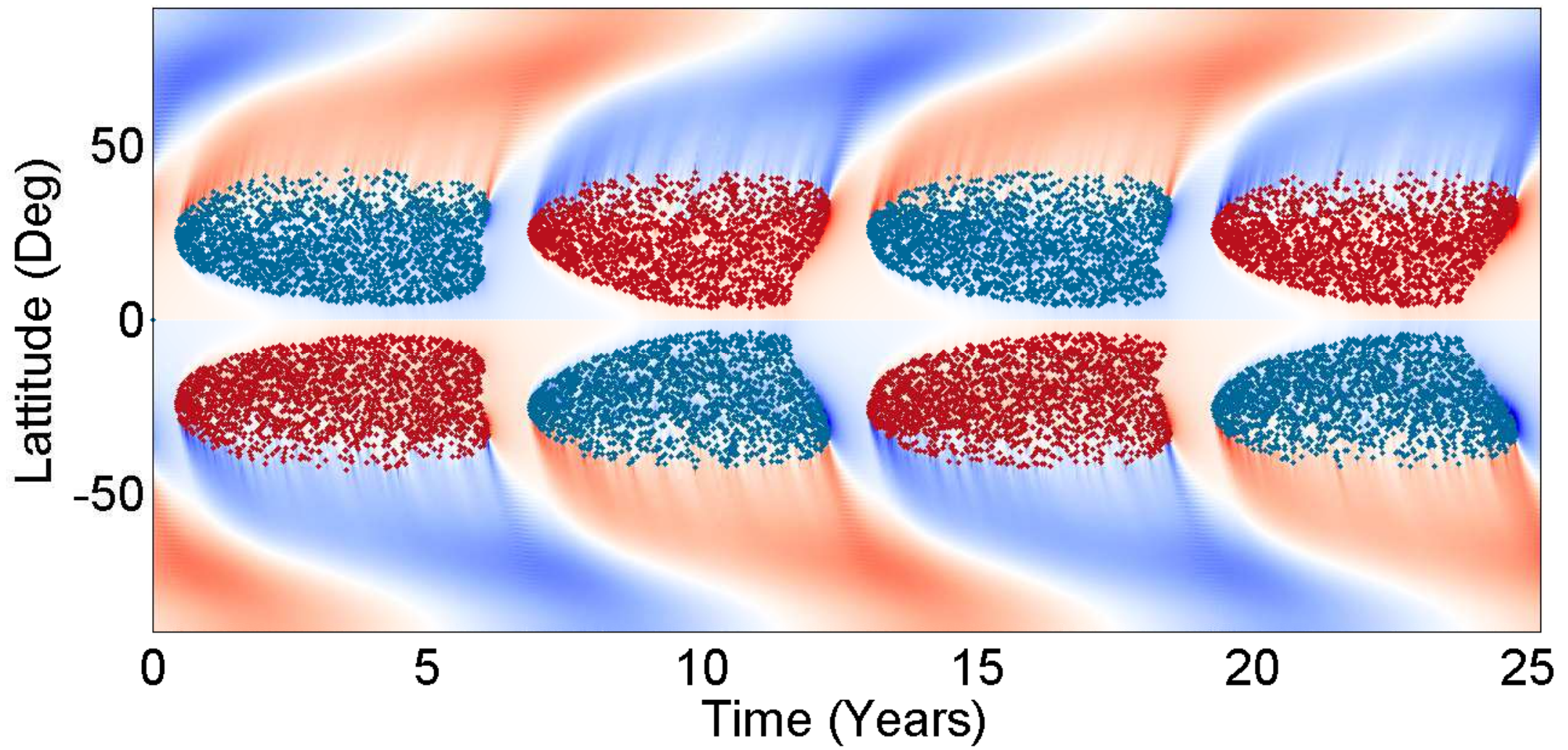} \\
                 (b)
  \end{tabular}
\end{center}
\caption{Synoptic maps (butterfly diagrams) showing the time evolution of the magnetic field in a simulation using the Mixing-Length Theory (MLT) estimate and diffusivity quenching (a), and a kinematic simulation using the geometric spatiotemporal average of the dynamically quenched diffusivity (b).  They are obtained by combining the surface radial field and active region emergence pattern.  For diffuse color, red (blue) corresponds to positive (negative) radial field at the surface.  The each red (blue) dot corresponds to an active region emergence whose leading polarity has positive (negative) flux.  We can see that a kinematic dynamo simulation in which we leave all parameters the same, but fix the diffusivity to the geometric spatiotemporal average, can capture the most important features of the magnetic cycle produced by the simulation using the MLT estimate and diffusivity quenching (period, amplitude and phase). }\label{Fig_5_Bfly}
\end{figure}

In order to compare the general properties of both simulations we cast the results in the shape of synoptic maps (also known as butterfly diagrams) as can be seen in Figure~\ref{Fig_5_Bfly}.  The results obtained using the MLT estimate and diffusivity quenching (Fig.~\ref{Fig_5_Bfly}-a) and the results obtained using a kinematic simulation with the geometric average fit (Fig.~\ref{Fig_5_Bfly}-c) are remarkably similar given the very different nature of the two simulations.  It is clear that the shape of the solutions differs mainly in the active region emergence pattern.  However, the general properties of the cycle (amplitude, period and phase) are successfully captured by the geometric average and are essentially the same.  This result argues in favor of the capacity of kinematic diffusivity profiles of capturing the essence of turbulent magnetic quenching.  These results define a framework which can be used to find a physically meaningful diffusivity profile based upon fundamental theory and models of the solar interior, rather than through heuristic approaches.

\subsection{Comparison with Solar Cycle Observations}

Ultimately, the goal of dynamo models is to understand the solar magnetic cycle and reproduce and predict it's main characteristics.  It's therefore important to compare our results with solar observations.  It is clear that the solutions are not exactly similar to those of kinematic dynamo simulations whose parameters have been finely tuned: cycle period of 7 years instead of 11, broad wings and incorrect phase.  This differences point to an overestimation of the turbulent diffusivity; mainly near the surface (affecting phase and period) and at the bottom of the SCZ (which affects period and the shape of the wings).  The cause of this overestimation resides in our definition of diffusivity quenching: in this work we use the average kinetic energy present in convection, which means that diffusivity is quenched equally through the convection zone.  However, convection is less energetic near the bottom of the SCZ (due to low convective speeds) and near the surface (due to low mass density).  This means that simulations taking this factor into account will probably yield more correct solutions.  Nevertheless, the solutions we obtain are reasonably accurate, given the fact that we have completely refrained from doing any fine tuning, instead limiting ourselves to fundamental theories and models.

\section{Conclusions}

In summary, we have shown that coupling magnetic quenching of turbulent diffusivity with the estimated profile from mixing length theory, allows kinematic dynamo simulations to produce solar-like magnetic cycles, which was not achieved before. Therefore, we have reconciled mixing length theory estimates of turbulent diffusivity with kinematic dynamo models of the solar cycle. Additionally, we have demonstrated that kinematic simulations using a prescribed diffusivity profile based on the geometric average of the dynamically quenched turbulent diffusivity, are able to reproduce the most important cycle characteristics (amplitude, period and phase) of the non-kinematic simulations.  Incidentally, this radial profile can be described by a double step profile, which has been used extensively in recent solar dynamo simulations. From the simulations reported here we provide an analytic fit to this double-step diffusivity profile that best captures the effect of magnetic quenching. A posteriori, our results strongly support the use of kinematic dynamo simulations as tools for exploring the origin and variability of solar magnetic cycles.

\acknowledgements

\section{Acknowledgements}

We are grateful to Dana Longcope and Paul Charbonneau for vital discussions during the development of this work.  We thank J{\o}rgen Christensen-Dalsgaard for providing us with Solar Model S data and useful council regarding its use.  The computations required for this work were performed using the resources of Montana State University and the Harvard-Smithsonian Center for Astrophysics.  We thank Keiji Yoshimura at MSU, and Alisdair Davey and Henry (Trae) Winter at the CfA for much needed technical support.  This research was funded by NASA Living With a Star grant NNG05GE47G and has made extensive use of NASA's Astrophysics Data System. D.N. acknowledges support from the Government of India through the Ramanujan Fellowship.

\bibliographystyle{apj}


\end{document}